Legerdemain in Mathematical Physics: Structure, "Tricks," and Lacunae in Derivations of the Partition Function of the Two-Dimensional Ising Model and in Proofs of The Stability of Matter

Martin H. Krieger
University of Southern California, Los Angeles CA 90089-0626

Abstract

We review various derivations of the partition function of the two-dimensional Ising Model of ferromagnetism and proofs of the stability of matter, paying attention to passages where there would appear to be a lacuna between steps or where the structure of the argument is not so straightforward. Authors cannot include all the intermediate steps, but sometimes most readers and especially students will be mystified by such a transition. Moreover, careful consideration of such lacunae points to interesting physics and not only mathematical technology.

Also, when reading the original papers, the structure of the physics argument may be buried by the technical moves. Improvements in the derivations, in subsequent papers by others, may well be clearer and more motivated. But, there is remarkably little written and published about how to read some of the original papers, and the subsequent ones, yet students and their teachers would often benefit from such guidance. I should note that much of the discussion below will benefit from having those papers in front of you—for which see note 1 below.

Legerdemain in Mathematical Physics: Structure, "Tricks," and Lacunae in Derivations of the Partition Function of the Two-Dimensional Ising Model and in Proofs of The Stability of Matter

Many a scholarly paper would seem to magically go from one line to the next, the reader not able to figure out the logic of the transition. Such legerdemain, whether it be magical or in doing physics, is no less impressive if you know how it is done, for you yourself would have to train extensively to actually perform these sleights of hand. What others ignored turns out to have needed informed careful attention. Yet, to be struck by legerdemain you must have actually read the paper, so that the device or method would stop you cold. Where did that come from? How do you get from line $A$ to to line $A+1$?

What Is Mathematical Physics?
Mathematical physics employs rigorous mathematics to develop physical theories and to solve physical problems. Those solutions may not be exact, and only provide approximations (but with rigorously derived errors). In any case, in the process of doing the mathematical work, one often discovers conditions needed to make the mathematics do the needed work and those conditions often are of physical significance, so that pointwise vs. uniform continuity of a series may, for example, allow a proof of a phase transition or not.

On the other hand, one may provide an exact solution, but not having done a careful job with your deltas and epsilons, or with your claim of analytic continuation, the solution is not mathematically rigorous. (Baxter's exactly solved models are not derived rigorously. In Kaufman and Onsager's, 1949, solution for the spontaneous magnetization, they did not publish the results since Onsager did not know how "to fill out the holes in the mathematics—the epsilons and deltas" concerning limits of Toeplitz matrices.[1] But Onsager did reveal his answer.)

Examples
Here I shall be concerned with two examples of mathematical physics. Solutions for the partition function of the two-dimensional Ising model and proofs of the stability of matter.

**The Ising Model**
1. The Ising partition function, numerically
2. Yang's plane rotation
3. Schultz, Mattis, and Lieb's particles
4. Wu, McCoy, Tracy, and Barouch's ending up with $P_{III}$
5. Baxter's working
6. Wilson's simplifications
7. Onsager's original paper;
**The Stability of Matter**
8. Dyson and Lenard's "hacking through a forest of inequalities"
9. Lieb and Thirring's wonderful approximation
10. Fefferman's "gruesome" calculations

In reading a paper in mathematical physics, it is not unusual to find that the transition from line $A$ to line $A+1$ is not clear or obvious, at least from what the author says. Eventually, you figure it out, or trust the author so you can get on with the rest of the argument or proof.

Or, a curious and surprising object is defined, with perhaps little motivation provided at that point. As for the object, its effectiveness may be revealed further down in the paper, and perhaps you have seen it employed elsewhere by that author or in other work.

Structure

Moreover, the structure of the argument or proof may seem so unmotivated, except that it does the work, and you wonder where it came from. Often, one suspects that that structure is discovered in the proving, and then the proof is presented is a much more lovely fashion than the author originally followed, yet the meaning of that structure, other than it serves the exposition, is elusive.

As for the *structure of the argument*, I have elsewhere (see my *Doing Mathematics*, edn. 2 (*DM*) and *The Constitutions of Matter* (*CM*)): described the structure of some of these papers.[2]

In these cases I have been concerned with the overall structure of the papers, rather than the line-by-line transitions and the legerdemain at various points that allows the paper to actually work. Usually, the authors tell you about the structure, early on in the paper, so you can follow their argument, but what they tell may not be explicit enough, nor accompanied with a diagram, say, so that you actually see that structure.[3]

What might seem like an argument driven by line-by-line inference is actually part of a strategy and structure that not only makes strategic sense, it also makes physical sense. That may be a matter of realizing the essential physics of the problem, or perhaps the proving leads to the author's realizing the essential physics of the problem and how to present the proof in a strategic fashion.

**The Two-Dimensional Ising Model**

The *Ising Model* of a ferromagnet is a two-dimensional grid of magnetic spins, spins that would prefer to be aligned: the Hamiltonian is

$-[E_1\sigma_{i,j}\sigma_{i+1,j} + E_2\sigma_{i,j}\sigma_{i,j+1}]$ summed over all spins and all configurations, where $\sigma = \pm 1$;

but if the temperature is above a critical point, the random motions due to thermal energy impede that. Computing the partition function or sum over states is a problem in classical statistical mechanics, the technical problem being to enumerate the possible states and their energies in as automatic a way as possible: algebraically, combinatorially, symmetrically, or graph-theoretically.

A. The Meaning Of The Numbers

It is sometimes useful to get into the inside baseball, even when we have fine formula solutions to problems. By inside baseball I mean the quantitative, numerical, details—and how they might be illuminating.

Lee and Yang, 1952, proved that at the critical point, zero magnetic field, in the infinite volume limit, the grand partition function of statistical mechanics (1/[exponential of the free energy]) has a zero. Put differently, there is a state of effectively zero energy, what we might expect were we to have an infinite specific heat due to an infinity of zero energy degrees of freedom. But to see the approach to that zero, one had to either compute the relevant part of the grand partition function for a finite number of particles, as the number of particles go to "infinity," or that same part of the GPF at the infinite particle level for values of the interaction near the critical point.

There is a very different sort of lacuna, one that hides legerdemain, that is achieved so-to-speak mechanically. We'll examine exact solutions to the Ising model in two-dimensions, for ln*PF*, and correct numerical calculations for *N* spins (or rows), as sums and as approximated by integrals for an infinite number of particles. We go into some detail to uncover what is going on in these calculations, their actual physics rather than their merely being formal or mathematical.

1. The two parts of the differing formulas, see below, for the ln*PF*, split into *a constant* and *a sum*, the sum representing a high-degree *polynomial*.
2. Yet they all must come to the same number, and so we see how *one addend makes up for the other* (except in Baxter, where they are not obviously separated).
3. The different formulas have *different ways of adding up the relevant physical objects that make up the Ising lattice, the quasiparticles' energies*. Onsager and McCoy/Wu have the same objects it would seem (albeit of different energy scales).
4. The formulas actually produce the *correct high and low temperature values for the* ln*PF*, as we might hope.
5. *The numerical approximations are in fact physical*, taking into account more and more spins in the Ising lattice. The sums are sums of quasiparticle free-energies.

6. Lee and Yang teaches us that *the partition function should have a zero at the critical point*, but as they point out that is only seen in the polynomial factor, and only in the infinite-$N$ limit. For it is only then that a monic polynomial with all positive coefficients could have a real zero.

The formulas below are for the ln$PF$, the logarithm of the partition function for the two-dimensional Ising model with the same coupling horizontally and vertically, $K$. At the critical point $K_c$=0.440687… (sinh$^2 2K_c = 1$). The first formula is from Baxter, the third is from Onsager, and the fifth is from McCoy/Wu.[4] One and Three are *exact* to the relevant sums for N spins (or rows), which the authors necessarily provide, for those sums are what they actually calculate. The integrals (2, 4, 5) are for an infinite number of particles. Note that $\exp \int_{0 to \pi/2} \ln$ is a geometric mean of the integrand.

(1/2) $\sum_{j,1 \text{ to } 2p}$ ln(2 cosh$^2 2K$ + 1/$k$(1+$k^2$-2$k$ cos($\pi(j$-1/2)/2$p$))$^{1/2}$), where $k$=1/sinh$^2 K$;   Baxter 1

(1/2$\pi$) $\int_{0 to \pi}$ F($\omega$) d$\omega$= (1/2$\pi$) $\int_{0 to \pi}$ ln(2 cosh$^2 2K$ + 1/$k$(1+$k^2$-2$k$cos 2$\omega$)$^{1/2}$) d$\omega$, where $k$=1/sinh$^2 K$;

Baxter 2

$n$/2 ln 2sinh2$K$ + (1/2) $\sum_{r,1 \text{ to } n}$ [acosh(cosh 2$K$ cosh 2$K$*-cos ($\pi(2r$-1)/2$n$)]         Onsager 3

½ ln 2sinh2$K$ + (1/2$\pi$)$\int_{0 to \pi}$ {$\gamma(\omega)$= ([acosh (cosh 2$K$ cosh 2$K$*-cos $\omega$]} d$\omega$,
where sinh 2$K$*=1/sinh 2$K$                                                                Onsager 4

ln($\sqrt{2}$ cosh 2$K$) +(1/$\pi$) $\int_{0 to \pi/2}$ ln(1+$\sqrt{(1- \kappa^2 \sin^2 \omega)}$) d$\omega$, where $\kappa$=2 sinh 2$K$/cosh$^2$ 2$K$

McCoy/Wu  5

The integrand for the Onsager formula may be seen to be <u>the energies (in $k_B T$ units) of the individual particles</u> that make up the lattice. (Onsager calls the integrand $\gamma(\omega)$, referring to an angle $\omega$ in a hyperbolic triangle, whose sides are 2$K$ and 2$K$*, $\gamma(\omega)$ being the third side and opposite the angle $\omega$. He then explains the integrand's behavior in terms of the triangle.) Schultz, Mattis, and Lieb indicate that these $\gamma(\omega)$ are energies of quasi-particles, in effect orderly rows. Hurst points out that, in particle physics, at the singular point, a meeting of two singularities, there is the threshold for the appearance of particles. *At the critical point the quasi-particles' energies begin at 0*, allowing for an infinite specific heat.

We might ask what is the meaning of the summands or integrands for the Baxter and the McCoy/Wu formulas? We note that cosh$^{-1}(x)$=ln($x$+($x^2$-1)$^{1/2}$). The Baxter and McCoy/Wu integrands or summands are of the form ln($A$+$B^{1/2}$) (I am assuming that they can be made to fit cosh$^{-1}(x)$=ln($x$+($x^2$-1)$^{1/2}$), so they echo the Onsager summand. Moreover, numerically, here are the values of the integrands as a function of $\omega$ *at the critical point* for each formula: Baxter, Onsager, McCoy/Wu.

<u>These are all *at the critical point*.</u>

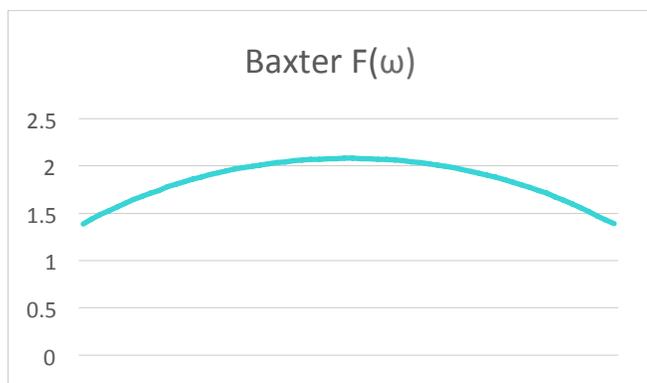

Baxter F($\omega$)

↑ Baxter Formula, ω: 0→π

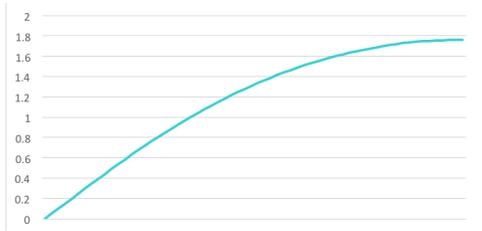

↑ Onsager Formula for γ(ω), ω: 0→π

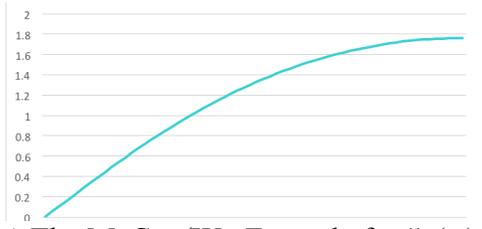

↑ The McCoy/Wu Formula for "γ(ω)," ω: 0→π/2

By eye, the $(1/2\pi)\times$Baxter area is about 0.95, $(1/2\pi)\times$Onsager is about 0.5+; and $(1/\pi)\times$McCoy/Wu is about 0.25, where the length of the axis is π, π, and π/2, respectively. When the constant term is added in, all should come to the same value for lnPF. We expect that ln PF at the critical point ($K_c$=.440687) to be about 0.93 (0.929695288=ln 2.533737= ln($\sqrt{2} \times \exp(2G/\pi)$)), where $G$ is Catalan's constant.

As for the numerical integration:

|  | Constant | Divide integral by | →to get | and the Σ is |
|---|---|---|---|---|
| Onsager | 0.347 | 2π→ . | .59 | 0.93 |
| McCoy/Wu | 0.693 | π→ | .23 | 0.92 |
| Baxter[5] | 0.69 | 2π→ | .23 | 0.92 |

Note, again, the difference in the Onsager and McCoy/Wu sums, nicely compensated for by the difference in the constants.

If we plot the ln of the PF vs. $K$, we expect ln2 (=0.69) for small $K$, and $2K$ for large $K$, high and low temperatures respectively.

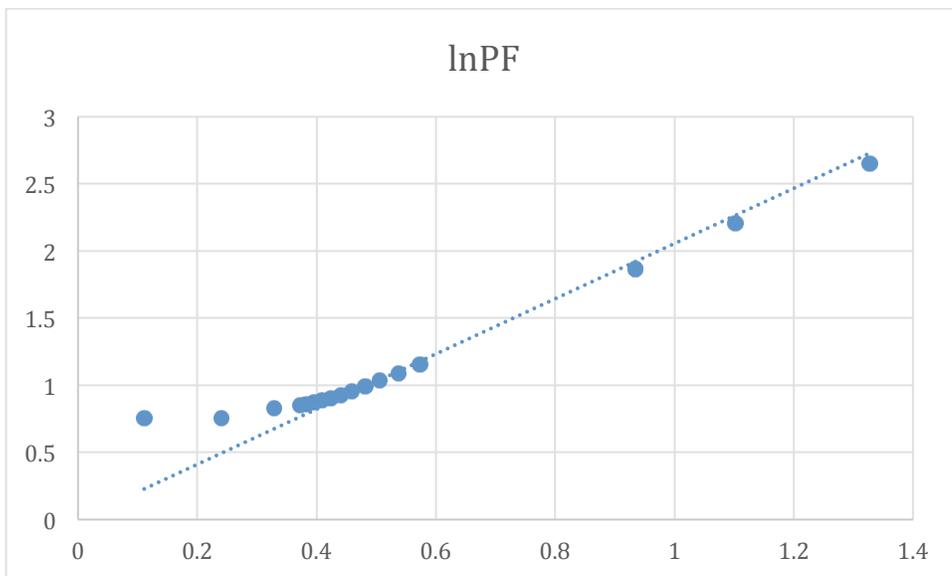

lnPF

For the chart below, I computed the sums, *using the exact N formulas* (*not* doing a "numerical integration") for 61 or 31 particles or steps for π and π/2. For the Onsager case, here is a chart (below) of the value of the numerical integration using varying numbers of steps. (For example, for N=4, we sample at π/8, 3π/8, 5π/8, 7π/8.) Even for four steps or particles we get a roughly-correct result. And at 25 steps and beyond things settle down. Since the number of steps is the same as the number of spins considered, we can see how the sum per spin gets smaller as we go to the infinite volume or N limit. We might expect this from the Lee-Yang theorem about the zeros of the grand partition function at the critical point as we go to infinite numbers of particles. We are adding up positive numbers the zero is approached only in that infinite particle limit.

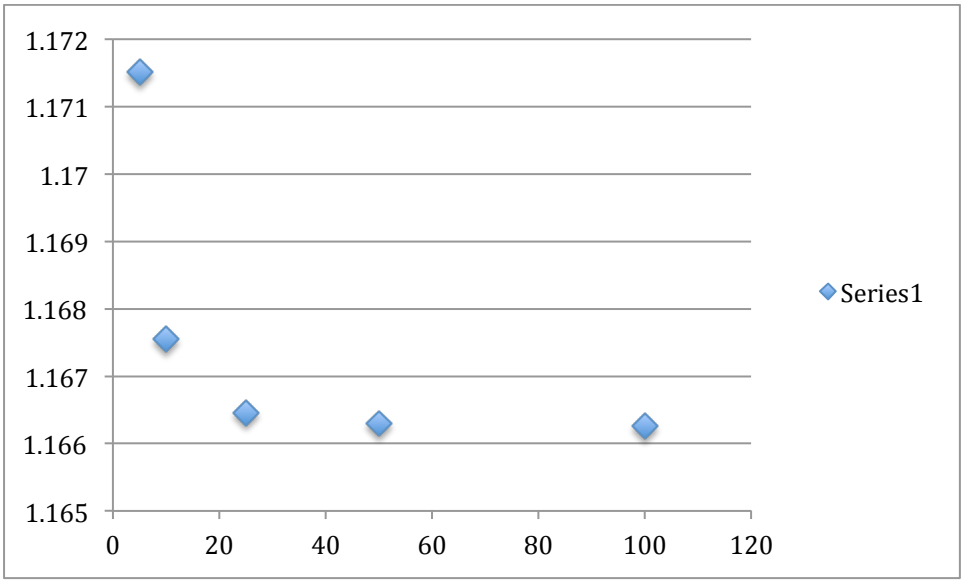

Onsager sum÷Number of Steps (or Spins or Rows) vs. Number of Steps (or Spins or Rows), at $K_c$

Now we might do a more conventional Monte Carlo integration. If we let the angle be randomly chosen, say with 100 random choices for ω, and average over say 14 trials, we get an estimate that is close to the expected one. But the variation is not so small, ±1%, and the average here is about ½% too small but well within the standard deviation.

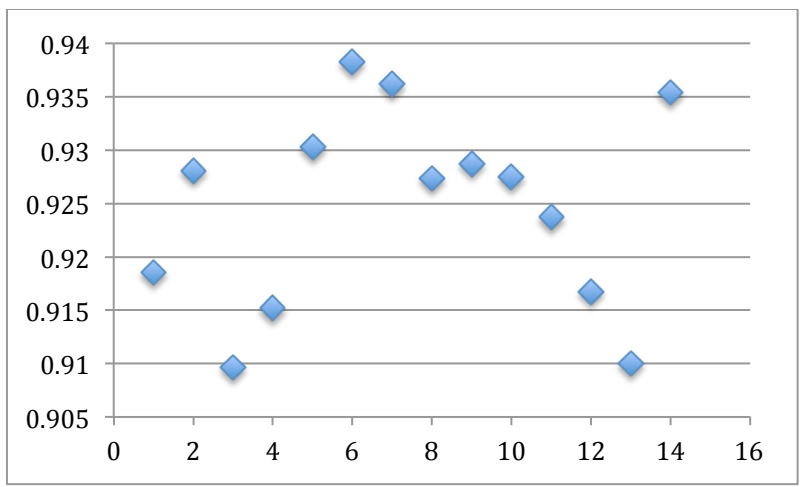

.
Value of lnPF for each each of 14 samplings, 100-random-tries for ω or ω. Average and SD: 0.92469652 and 0.009363738.

Again, Lee and Yang (1952) point out that at the critical point the partition function should exhibit a zero, in the infinite volume limit. Namely, one might think of the PF as being a Constant$^N$× Polynomial (here in tanh $K$), its degree being the number of particles or spins, $N$. Basically, in the partition function we have products of terms (cosh $K$+ σσ'sinh $K$) = cosh $K$ (1+ σσ'tanh $K$), and so we have a polynomial from the product of the (1+ σσ'tanh $K$) terms. (In other situations, one wants to avoid computing exponentials, and ($e^K$+ σσ'$e^{-K}$) becomes $e^K$(1+ σσ'$e^{-2K}$), and defining $K^*$ by sinh2$K^*$=1/sinh2$K$, we again get a (1+ σσ'tanh$K^*$) term

    Isolating the integral (now $N$ is infinite) in McCoy/Wu, 1973, and in Onsager, we get a good handle on the polynomial, the additional term ln(cosh2$K$) or 1/2ln(sinh2$K$) coming from the creation of the (1+ σσ'tanh $K^*$) terms. Below are plots of the sums for McCoy/Wu and for Onsager as a function of $K$. They have minima at the critical value of $K$, $K_c$=~0.44, noting again that we are dealing with ln$PF$.

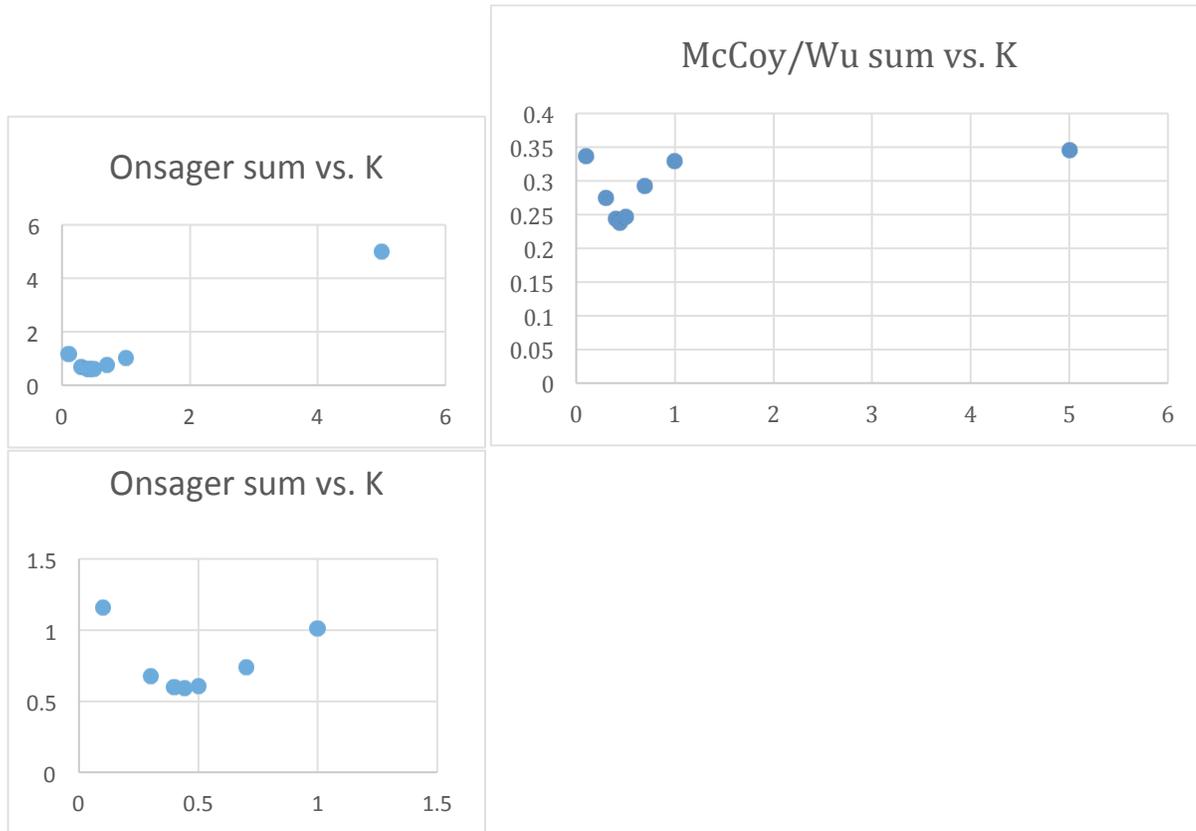

    Another approach might be: If we examine the above formulas for the ln$PF$, to get the sum it makes sense to compute ln$PF$-ln√sinh2$K$ (or -ln cosh$K$). Below we have shown (ln(PF/cosh$K$))^10, for k=0.8, 0.9, 1.0, 1.1, 1.2, 1.3. At the critical value, k=1.0, there is a minimum.

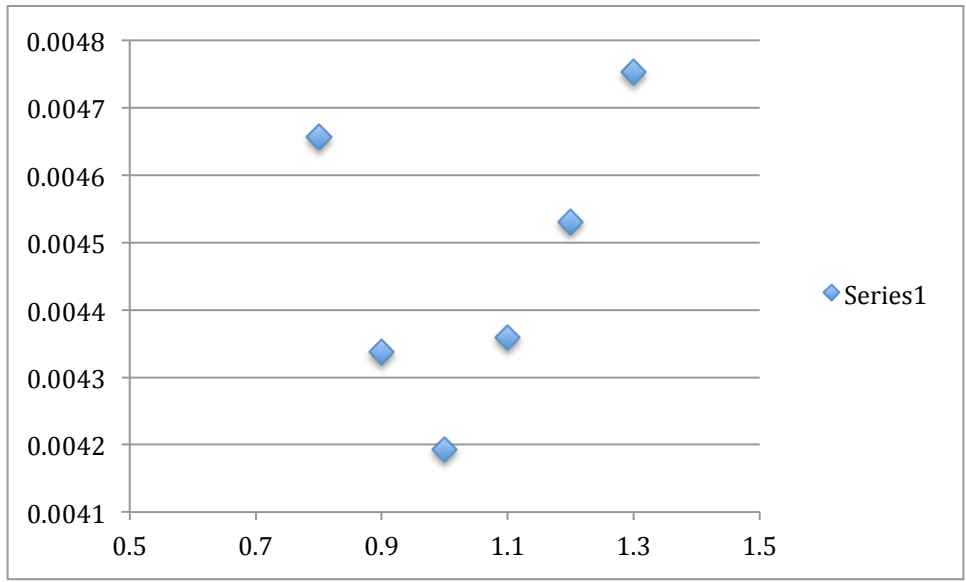

B. An Amazing Invention
An amazing invention of Yang, 1952, appears in the middle of an ingenious and swerving derivation of the spontaneous magnetization of the Ising model in two dimensions. Onsager's paper was generally known to use methods unfamiliar to most physicists, and his student Kaufman's reformulation in the more familiar terms of spinors enabled Yang to do his work.2. *We notice that ...*
   Yang, 1952, at one point in his derivation of the spontaneous magnetization of the Ising model needs to rotate a matrix so as to diagonalize it. What he does is quite inventive, a way forward that is not at all obvious. Later derivations perhaps make his move a bit more understandable, for which see the note.[6] He says that,

> $1+C_n$ does not induce a rotation. <u>But we notice that</u>
> $1+C_n = \lim_{a \to i\infty} \cos^{-1} a\,(\cos a - i\,C_n \sin a) =$
> $\lim_{a \to i\infty} \cos^{-1} a\, \exp(-iaC_n)$  (22)  and $\exp(-iaC_n)$ does induce a rotation. (p. 810)

Yang says elsewhere that this was the longest calculation in his career. His noticing-that, before it was realized, was likely one of the slowdowns in his calculation. (He was not so worried about rigor, it would appear.) He employed "we notice that" again later in the paper.

C. Employing a Trick of the Past
Lieb, Mattis, and Schultz, 1961, studying one-dimensional models in mathematical physics, had employed a transformation that turned mixed fermion/boson operators into fermion operators, a transformation attributed to Jordan and Wigner, 1928. In their redoing (1964) of Onsager's algebraic derivation, using more familiar notions of second quantization and diagonalizing a quadratic form, they found, instead of Onsager's hyperbolic triangle, particles and (Cooper) pairs of particles. They could explain Onsager's integrand as particle energies rather than as the side of a hyperbolic triangle.
   Schultz, Mattis, and Lieb, 1964, in their algebraic and rather more familiar to contemporary physicists derivation of the Ising model, are confronted with operators, the Pauli spin operators, $\sigma^{x,y}_i$, that anticommute at a site, but for different sites the spin operators commute. (p. 861)

> <u>It is nevertheless well-known</u> how to change such operators to ones obeying a complete set of anticommutation rules [fermions]. We introduce annihilation and creation operators by the formation $C_m = [\exp(\pi i \Sigma_{1 \text{ to } m-1} \sigma_j^+ \sigma_j^-)] \sigma_m^-$ [obviously not at all local as an operator] a method Lieb and collaborators had already employed in earlier work, and was attributed to Jordan and Wigner (1928).
   Further along in their derivation, with the $C_m$'s now Fourier transformed to $\eta_q$'s they find themselves with expressions such as $\eta^\dagger_q \times \eta_q$.
   The paired nature of the quadratic forms…are <u>extremely reminiscent of</u> the pair-Hamiltonian and the energy states introduced by Bardeen, Cooper, and Schrieffer in the theory of superconductivity, particularly in Anderson's formulation... We can simplify the results for $V_q$ [which is written in terms such as $\eta^\dagger_q \times \eta_q$] in precisely the same way Bogulubov and Valatin simplified the BCS theory, if we introduce the transformation

$\xi_q = (\cos \varphi_q) \eta_q + (\sin \varphi_q) \eta^\dagger_{-q}$

The particle energies $\varepsilon_q$ ($=\gamma(\omega)$) are those of the $\xi_q$ and $\xi_{-q}$ and the (Cooper) pairs, designated by $b^+$ in SML are $\xi_q \xi_{-q}$. That is, not only is their method more familiar, but they interpret their results in contemporary terms.

   Still, the algebraic solutions of Onsager and of Kaufman were not intuitive for many physicists. …[There is] the considerable formidability of the method [employed by Onsager, 1944, to solve the two-dimensional Ising Model]. The simplification by Bruria Kaufman using the theory of spinor representations has diminished, but not removed, the reputation of the Onsager approach for incomprehensibility, while the subsequent application of this method by Yang to calculate the spontaneous magnetization has, if anything, restored this reputation.
   …[the present paper presents] the algebraic approach in a way that is both very simple and intimately connected with the problem of a soluble many-fermion system. Except for one or two crucial steps, the approach is straightforward and requires no more than a knowledge of the elementary properties of spin ½ and the second quantization formalism for fermions. … The two-dimensional Ising model, rather than being entirely different from the trivially soluble many-body problems, reduces in some ways to one of them, being just the diagonalization of a quadratic form. [Schultz,Mattis,Lieb, p. 856]

D. Where Did That Come From?
   Even more strenuous and hairy and lengthy were Wu, McCoy, Tracy, and Barouch's, 1976, account of the correlations of spins when very close to the critical point and for very large lattices (the scaling regime). The length and complexity of the paper is striking, ending with a distant cousin of the trigonometric functions (for example, already earlier discussed in Ince's textbook *Ordinary Differential Equations*, 1926, pp. 345ff). I have no idea of the source of their persistence and stamina, although I am told that patience is crucial in this sort of work. The authors are a sequence of descendants of an expert on antennas and electromagnetic theory, Ronold King, and presumably Wu learned his skill from his teacher, one of whose other students, in working on antenna theory, had concluded his dissertation work with the same distant cousin of the trigonometric functions. Following Wu and collaborators' earlier detailed calculations, years later, Wu, McCoy, Tracy, and Barouch, 1976, after a long and complicated calculation, found that the spin-spin correlation in the scaling regime of the Ising model may be expressed using one of the <u>Painlevé transcendents</u>, $P_{III}$, distant cousins of the sines and cosines. As far as I can tell that was not at all expected until WMTB arrived there. But it is worth noting that <u>Wu is a student of Ronold King, whose student Myers encountered an integral similar</u> to the one found by WMTB, in studying scattering of electromagnetic waves from a strip. The integral is $\int_{a \text{ to } b} K_0(|x-x'|) \varphi(t) \, dt$ ($K_0$ being a Bessel function), solvable by Wiener-Hopf techniques, techniques Wu employed elsewhere in his work on Ising. The genealogy is:

King(PhD 1932, Harvard)→Wu(1956)→McCoy(1967)→Tracy(1973, StonyBrook), ?McCoy advised Barouch (1969) at Stony Brook.

→Myers (1963)

There is a sequence of work (Domb et al) of numerically solving the Ising model using *n*-particle cluster expansions, especially to compute scaling exponents. But the next step was to appreciate a deep symmetry of the system, as did Wilson (following Kadanoff and others) in his work on the renormalization group.

E. "A Useful Identity, Easily Seen"

The impetus for writing about legerdemain was a blog post that explained the transition from Equation 106 to Equation 108 in Onsager, 1944, the exact solution of the Ising model, supplying as well a proof of Equation 107, a formula needed for the transition.

In Lars Onsager's (1944) derivation of the exact partition function for the two-dimensional Ising Model, there are moments when the transition between lines may need a bit more elaboration, not to speak of his use of Lie algebra. So for example, his formula (106) for the partition function, PF (his $\lambda$), in the limit of an infinite number of particles, might be shown to be symmetrical between the horizontal and vertical couplings, *K* and *L* (rather than Onsager's *H* and *H'*), of the spins on the lattice.

(106) $\ln PF = 1/2 (\ln 2 \sinh 2K) + (1/2\pi) \int_{0 \text{ to } \pi} \gamma(\omega) d\omega$

where $\cosh \gamma(\omega) = \cosh 2L \cosh 2K^* - \sinh 2L \sinh 2K^* \cos \omega$, and
where $K^*$ is defined by $\sinh 2K \sinh 2K^* = 1$

He says,
> There are several ways to show that (106) actually describes a symmetrical function of *K* and *L*. For example, with aid of the <u>useful identity</u>
>
> (107) $\int_{0 \text{ to } 2\pi} \ln (2 \cosh x - 2 \cos \omega) d\omega = 2 \pi x$
>
> we can convert (106) into the double integral (108) [the symmetrical form]… ) It seems rather likely that this result could be derived from direct algebraic and topological considerations without recourse to the operator method used in the present work. (pp. 248-249)

[(108) $\ln (PF/2) = 1/(2\pi^2)$
$\times \int_{0 \text{ to } \pi}) \int_{0 \text{ to } \pi} (\ln (\cosh 2K \cosh 2L - \sinh 2K \cos \omega - \sinh 2L \cos \omega') d\omega d\omega']$

In all subsequent discussion and derivations of the work, it seems that nowhere is (107) discussed, or is the derivation of (108) from (106) using (107) shown. Now, other derivations (except for Baxter's) get the double integral form directly, so they need not address the issue. (They often then go to the single integral formula.)

COV (see note 6) provides the details in a blog post, both the proof of (107) and the derivation of (108). As for proving (107) that will involve a statement about the geometric mean of $(1 + r \exp i\omega)$. And then for (108), from (106) using (107), where *x* is now $\gamma(\omega)$, and so $\cosh x$ (that is, $\cosh \gamma(\omega)$) can now be under its own integral in (106), and so eventually we have (108). (Note that if *K=L*, $\cos \omega + \cos \omega'$ may be transformed into $\cos \omega \times \cos \omega'$ making for a nicer integral. (That is, $\cos (\omega+\omega') = \cos \omega \cos \omega' + \sin \omega \sin \omega'$, and then doing the same for $\omega-\omega'$, and so that $\omega$ becomes $(\omega+\omega')$.)[7]

F. Signposting the Way
Rodney Baxter, 1982, made ingenious use of the symmetries of the Ising lattice, functional equations, properties of elliptic functions in the complex plane (used here to make square roots single valued, "uniformized") to exactly solve many lattice models. Again, there appears legerdemain, a rather different employment of those elliptic functions Onsager used, and convenient but unproved assumptions of analytic continuation. Early on, Baxter was influenced by Lieb.

Baxter's *Exactly Solved Models in Statistical Mechanics*, 1982, and his papers are filled with adept use the symmetries of the transfer matrix in terms of *K* and *L*, a variable *k* that is both a measure of temperature and the modulus of the elliptic functions (as in Onsager as well), functional equations, and assumptions about analyticity and analytic continuation.  (In a paper with Enting, he uses all of these, especially the star-triangle relation, plus a remarkable geometric reconception of the lattice itself, to derive functional equations for the partition function.) His approach allowed for generalization to many other models, and like Onsager he provides an Appendix on elliptic functions.

As for intermediate results, Baxter is direct and honest:

> I knew from experience how many sheets of paper go into the waste-paper basket after even a modest calculation: there was no way they could all appear in print.
> 
> I hope I have reached a reasonable compromise by <u>signposting the route</u> to be followed, without necessarily giving each step. …I discuss the functions k(α) and g(α) in some detail, since they provide a particularly clear example of how elliptic functions come into the working. . . .I merely quote the result for the spontaneous staggered polarization..., and refer the interested reader to the original paper: its calculation is long and technical, and will probably one day be superseded…" (p. v)

Fefferman, # 10 below, signposts using colloquial expressions and oversimplified abstracted versions of a calculation.

G. "Further Details Of Simplifications Like This Will Not Be Reported Here"

Kenneth Wilson, 1975, came to understand some of the mysteries of quantum field theory by actually calculating results numerically on a computer, employing simpler models still possessing the mystery. Actual numerical calculations (again computationally) allowed him to understand systems with an infinite degrees of freedom, not by radically truncating those degrees of freedom, but by finding the crucial degrees of freedom. The meaning of the formulas and models was found numerically, namely by which approximations actually worked. (Onsager guessed how to go about an exact solution by starting with lattices with 2, 3, and 4 infinite rows of spins, and solving them by hand.)

Wilson is reporting on his application of the renormalization group to solving the two-dimensional Ising mode, using an algorithm and a computer and approximations along the way. He says,

> The sum over all configurations of the old spins was carried out sequentially, …only the coupling of $s_{15}$ to nearby spins was included in the calculation. The "nearby spins" are those shown in Fig. 9.…In practice it was possible to reduce the nearby spins to the subset *C* of 13 spins shown in Fig. 9. …<u>Further details of simplifications like this will not be reported here.</u>

In the actual practice of programming and discovering which interactions might be ignored (for they had little actual effect on the numbers) the "details" become apparent, perhaps in retrospect made motivated and scientific. Just which terms in an expansion are retained, which left behind, is a craft skill. Wilson, it seems, has always been interested in approximations that allow you to solve problems and get out actual numbers. And computation and programming has been one of his methods, the other being perturbation theory in quantum field theory using Feynman diagrams.

There is another numerical fact—all the formulas for the partition functions are approximations for an infinite particle system, the actually computed formulas for sums to *N* spins being replaced by integrals.

**The Stability of Matter**
The *stability of matter* is something we take for granted, but a rigorous proof eluded all since the beginning of quantum mechanics. Namely, the ground state energy of a clump of matter, namely due to the Coulomb electrical forces among the electrons and nuclei, should have a lower bound *proportional to the number* of atoms or molecules in that clump (rather than say $N^2$). Onsager, 1939, had provided an argument that was rough but captured the main features. And, later, Ruelle revived the problem.[8]

It is striking how the different modes of solution provide very different perspectives on the Ising lattice or on matter's stability. Moreover, the mathematics serves the physics, albeit some of the time that service is not so apparent. By understanding how an author constructs the argument and the structure of a mathematical physics account, one is likely to discover more deeply the essential physics of the system. I suspect that some of the time, the author finds ways of moving forward in their argument, paying more attention to the mathematics than the physics, and only later, if ever, is the physics behind the devices and tricks revealed. Yet, that physics is almost always of great interest, displaying features of the system that were otherwise not appreciated.

H. "Hacking Through A Forest Of Inequalities"
In a hard-won calculation, Dyson and Lenard, 1967, proved the stability of matter. Their proportionality constant was about $10^{14}$, when we might expect it to be on the order of 1. Lenard interested Dyson in his problem, deriving from Lenard's work in plasma physics. Dyson's training as a classical analyst served him well, and their strategy put the calculation on the road. Retrospectively, he referred to it as "hacking through a forest of inequalities" (a usual theme in classical analysis).
    Again, the task is get an estimate of the ground state energy of a clump of matter, ignoring gravitation, and considering only the electrical forces between nuclei and electrons. We might expect an $N^2$-proportionality, given the number of interactions, but what we want is an $N$-proportionality. Dyson and Lenard, 1967, provided the first rigorous proof of the stability of matter, although Onsager had provided a non-rigorous but physically correct rough argument many years earlier. The Dyson/Lenard proof demanded a sequence of inequalities, each with a small proportionality constant, but the product of those constants was quite large, about $10^{14}$ in atomic units (1 Rydberg = 1/4 in atomic units). Dyson referred to their proof in terms of "hacking," but in fact the proof is physically well motivated, the headings of the sections being informative) albeit it does not have the enormous advantage of the physics provided in Lieb and Thirring.

I. "Thomas-Fermi Atoms Do Not Bind"
But it was Lieb/Thirring, 1975, who were then able to redo the proof, presumably suspecting that it could be done using a familiar model of an atom, the Thomas-Fermi model, familiar to Lieb from his earlier work with Simon, 1973. In a physically motivated way (namely, Thomas-Fermi atoms do not bind), rather more specific than Dyson/Lenard, they got a much better account of stability, not needing to "hack through a forest of inequalities" as had Dyson/Lenard. Lieb and collaborators improved on Dyson/Lenard, for the proportionality constant in the stability of matter, from $10^{14}$ to about 20, and then lower over the years.
    In Lieb and Thirring's, 1975, proof of the stability of matter, their constant is perhaps $10^{-12}$ th of Dyson-Lenard, largely because they get the physics right. Namely, Teller had shown that in the Thomas-Fermi model of an atom, there is no binding. So if TF atoms (with a different proportionality constant) could be shown to be a lower bound of the energy of matter, then the total energy would be proportional to $N$ atoms, and so stability would be proven. Lieb and Simon had earlier studied the Thomas- Fermi model.

J. "An elementary identity, fourier analysts are quite familiar with it. Gruesome details, nasty and ghastly calculations, modulo oversimplifications, applying general nonsense."
To give an account of atoms and of stability, Fefferman and collaborators, 1986-1997, employed rigorous classical mathematical analysis, to achieve a deeper understanding and better bounds on the binding energy/particle. Fefferman was a master of the devices of classical analysis; he had experience with "nasty" calculations and persistence—shown in his earlier work.

.
    Charles Fefferman says at one point:

    we can start with the <u>elementary identity</u>

$|x-x'|^{-1} = (1/\pi) \iint_{y \varepsilon \mathbf{R}^{\wedge}3,\, R>0} \chi_{x,\, x' \varepsilon B(y, R)}\, dy\, dR/R^5$  for $x, x' \varepsilon \mathbf{R}^3$ (25)

> …[except for $1/\pi$] identity (25) is forced by the fact that both sides transform in the same way under translations, rotations, and dilations." (p. 6, "On the Dirac…," *Advances in Mathematics*, 1994) where $\mathbf{R}^3$ is three-dimensional space, $y$ is the center of a Ball of radius $R$, $\chi$ is the indicator function for when $x$ and $x'$ are within the Ball.[9] Elsewhere he says of a related formula that "it is easily verified by using the Fourier transform. <u>Fourier analysts are quite familiar with it</u>…" p. S90 *Communications in Pure and Applied…*, 1986. He refers to the "<u>gruesome details</u>" of the calculation. (p. S88)

Fefferman constantly provides previews and motivations, often designating them in colloquial terms. In Fefferman's Bergman kernel paper, he speaks of "<u>nasty</u>" and "<u>ghastly</u>" calculations pp. 4 and 45; "<u>the idea is</u> to transfer the Bergman kernel" (9); "We apply <u>this general nonsense</u>…" (19), both meant to motivate and set up a complex calculation; "as a sum of terms <u>which (hopefully) transform to something recognizable</u> when expressed in ζ-coodinates" (37); "<u>Modulo oversimplifications</u>, this the plan of our proof of Theorem 1." (8), at the conclusion of a detailed motivating introduction.

Genealogy Reconsidered

One explanation of such legerdemain is to understand the genealogy of the authors and their previous work and teachers. So:

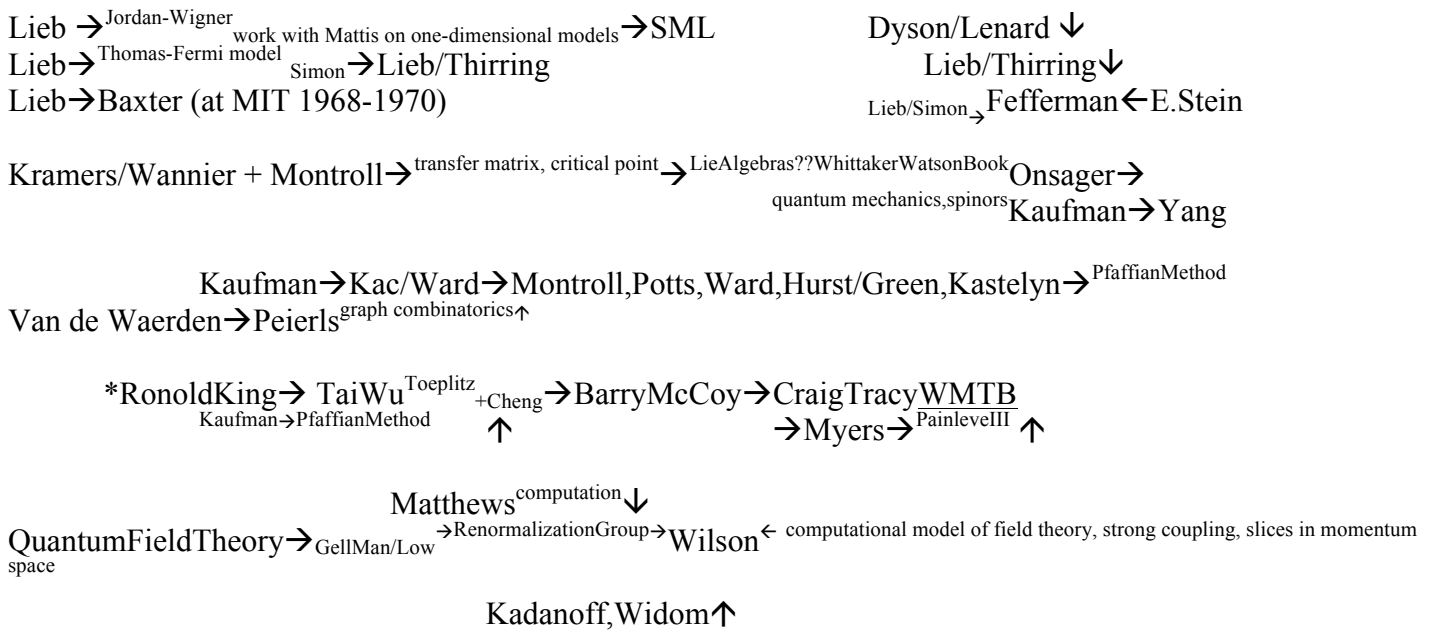

*KING WAS WU'S ADVISOR, WU MCCOY'S, MCCOY TRACY'S.

RONOLDKING→<sub>PHDHARVARD,1956</sub> WU→ <sub>PHDHARVARD,1967</sub>MCCOY→ <sub>PHDSTONYBROOK,1973</sub>TRACY
    RONOLDKING→<sub>PHDHARVARD,1963</sub> JOHNMYERS

**Legend:** SML=Schultz, Mattis and Lieb.  Other papers are given in the notes.[10]

**To Conclude**
In sum, legerdemain is a matter of what you happen to know, dogged persistence, ingenious invention apparently out of nowhere but likely from past work you have done, learning that something can be done so it makes sense to try to do it better using your repertoire of skills and knowledge, and the capacity for approximations to retain the crucial features of a system.

Acknowledgment: Professor JHH Perk pointed out an error on p.3, now corrected, for which I am grateful.

Notes

1. 1. See also the Baxter paper about the manuscripts (note #8 below).
   Onsager never published his derivation of this result, although in 1949 he and Kaufman produced a paper on the short-range order or, more correctly, on the set of pair-correlation functions of the square lattice Ising model. It was left to C. N. Yang to rederive the result independently. Only twenty years later, at the Battelle Symposium in Gstaad, did Onsager reveal fully that in computing the long-range order he had been led to a general consideration of Toplitz matrices and determinants but did not know how "to fill out the holes in the mathematics—the epsilons and the deltas." By the time he had done this, he found that "the mathematicians" had got there first—although, in fact, the generality and depth of Onsager's results were not matched for many years.
2. Onsager's 1944 Ising model exact solution (*CM* pp. 69-75, chs. 6, 7);
   Fefferman's (and Seco's) work, 1990-1996, on the ground state energy of atoms and of matter (*DM* 137-142, 165-182);
   Wu, McCoy, Tracy and Barouch on the scaling of the two-point spin-correlation function in Ising (*DM* 185-198-210) and Wu and collaborators' earlier work;
   Lieb and Thirring (*DM* 158-165), and Dyson and Lenard (*DM* 150-157), on the stability of matter;
   Yang on the spontaneous magnetization in Ising (*DM* 129-137);
   Baxter and Enting, 1978, on Ising (DM 179-180), as well as Baxter's more general strategy of going from symmetries of the lattice and its partition function to functional equations for the partition function.

3. See *Doing Mathematics,* Figs. 4.4-4.6, pp. 137-138; Figs. 4.7-4.9, 150-152; Fig. 4.13, 175; and, Fig 4.18, 197.
4. It would be interesting to show how the integrands are related, and in some papers the authors show how their solution corresponds to Onsager's original solution. McCoy/Wu actually get Onsager's solution for the double integral, and then derive this single integral. I do not believe Baxter tries to show equivalence. In the case of the Pfaffian and combinatorial solutions, as in McCoy/Wu, they show how the matrices of Kac and Ward are the same as theirs, and Kac and Ward show how their formula is like Onsager's.
5. If you modify the integral so that you have zero energy for Baxter, one gets

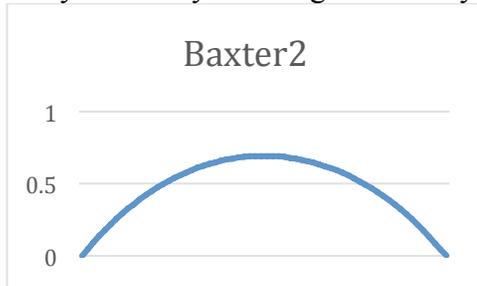

Recall that the base of the Baxter integrand was about 1.4, now it is zero.

6. See Yang, "The Spontaneous Magnetization," p. 811, Equation 34. Yamada suggests another perspective on the "artificial limiting process." He introduces another such process which then expresses the correlations as a ratio of determinants, but which determinants are zero in the limit (columns are equal). So he applies L'Hôpital's rule, and takes the derivatives of both numerator and denominator. But such a derivative, $d/d\omega$, is in effect an infinitesimal rotation. See, K. Yamada, "Pair Correlation Function in the Ising Square Lattice, Generalized Wronskian Form," *Progress of Theoretical Physics* **76** (1986): 602–612, at pp. 603, 608.
   Baxter claims to be following Yang in his more recent algebraic derivation. (R. J. Baxter, "Algebraic Reduction of the Ising Model, *Journal of Statistical Physics* **132** (2008): 959-982.) The artificial limiting process is given by Baxter's $\exp(-\gamma J)$ $\gamma \to \infty$ (this is not Onsager's $\gamma$), where $J$ is in effect the one-dimensional hamiltonian. The transfer matrix is represented by a Hamiltonian, $\exp-\alpha H$, that $H$ being much like Onsager's $A+k^{-1}B$. As for Kaufman and Onsager's work, Baxter has unearthed the original manuscripts. He describes the two ways they did the correlation function (Szegő limit theorem, integral equation), and provides a third way of his own invention. (R. J. Baxter, "Onsager and Kaufman's calculation of the spontaneous magnetization of the Ising model," *Journal of Statistical Physics* **145** (2011): 518-548; "Onsager and Kaufman's calculation of the spontaneous magnetization of the Ising model: II," *Journal of Statistical Physics* **149** (2012): 1164-1167; "Some comments on developments in exact solutions in statistical mechanics since 1944," *Journal of Statistical Mechanics: Theory and Experiment* (2010) P11037, 26pp.

Kadanoff indicates Yang's in-effect employment of Wiener-Hopf methods: Yang evaluated the trace of [an expression for the zero-field magnetization, Kadanoff's 3.30, see Figure 4.3 in our text] by finding all the eigenvalues of the operator inside the ln [of 3.30] and then performing the indicated summation. The calculation is possible because the difference equation that appears in the eigenvalue problem is solvable through the use of the Wiener-Hopf method. L.P. Kadanoff, "Spin-Spin Correlations in the Two-Dimensional Ising Model," *Il Nuovo Cimento* **B44** (1966): 276–305, especially pp. 289–293, at p. 292. Yang actually solves the equation that gives the eigenvalues "by inspection" in IIIC. He never needs to refer to Wiener-Hopf. (Note that Yang's $\Omega(z)$ at Equation 61 is Onsager's $\exp i\delta'$. Also, the product that appears becomes a sum when one takes the logarithm, as for the Szegő limit theorem.)

7. Then Onsager, using Kramer and Wannier's, 1942, variable $\kappa$ ($=2\sqrt{k}/(k+1)$, $k=1/(\sinh^2 2K)$, when $K=L$ (actually K+W's kappa is ¼ of the kappa I use) symmetric around the critical temperature, "expand[s] the logarithm [in 108] in powers of $\kappa$ and $\kappa'$ and integrate term by term, to get a series--which when $K$ and $L$ are the same, so $\kappa$ and $\kappa'$ are the same--in $\kappa^{2n}$ with binomial-like coefficients (109c): $\ln PF - \ln(2\cosh 2K) = \ln(1 - (1/16)\kappa^2 - (1/256)4\kappa^4 \ldots)$, at the critical point the $\ln(2\cosh 2K) = 1.04$, and $\ln(\text{series}) = -0.089$." Note that this kappa, too, is four times Onsager's kappa.

In his Appendix, using an expression $\Phi(u)$ whose integral is proportional to $\int_{0\text{to}\pi} \gamma(\omega) d\omega$, Onsager says, "It is easily seen that $\Phi(u)$ satisfies the standard conditions for development in partial fractions…" and then one might "readily perform[ed]" the integral of $\Phi(u)$ and so in effect $\int_{0\text{to}\pi} \gamma(\omega) d\omega$. (p. 258). "To obtain an expansion suitable for computation in the [that] region" of the critical point, one finds a version of $\Phi(u)$ that "is a periodic function of $u$." "Its Fourier series is easily derived with the aid of the identity …from those of the Jacobian elliptic functions," for which he gives a reference to Whittaker and Watson's *Modern Analysis*.

Earlier in the paper, Onsager uses the theory of group representations to diagonalize a pseudo-hamiltonian. Subsequently, Kaufman, 1949, redid Onsager, for a a finite lattice, using the apparatus of spinors and spin-representations of the orthogonal group, which made Onsager's method more recognizable to many physicists. B. Kaufman, "Crystal Statistics II: "Partition Function Evaluated by Spinor Analysis," *Physical Review* **76** (1949): 1232-1243.

8. L. Onsager, "Electrostatic Interaction of Molecules," *Journal of Physical Chemistry* **43** (1939): 189-196; D. Ruelle, "States of Classical Statistical Mechanics," *Journal of Mathematical Physics* **8** (1967): 1657-1668.

9. Let $V(x,y)$ be the integral over all $z$ in $\mathbf{R}^3$ and $r>0$ of the indicator function of the set where $x$ and $y$ both lie within distance $r$ of the point $z$, divided by $r^5$. Then one sees from the definition of $V(x,y)$ that $V(x,y) = V(x+w,y+w)$ for any vector w, therefore $V(x,y) = V(x-y,0)$. Moreover, $V(Tx,0) = V(x,0)$ for any rotation T of $\mathbf{R}^3$, as one also sees from the definition of $V(x,y)$. Therefore, $V(x,y)$ is a function of the distance from $x$ to $y$. Again from the definition of $V(x,y)$, one checks easily that $V(tx,ty) = V(x,y)/t$ for any positive real number t. Therefore $V(x,y)$ has the form constant/(distance from $x$ to $y$). Evaluating $V(x,y)$ for one particular pair of points, one finds that the constant is $\pi$. (as provided by CF to MK 3Ap18)

10. C. A. Hurst and H. S. Green, "New Solution of the Ising Problem from a Rectangular Lattice," *Journal of Chemical Physics* **33** (1960): 1059-1062; M. Kac and J. C. Ward, "A Combinatorial Solution to the Two-Dimensional Ising Model," *Physical Review* **88** (1952): 1332-1337; E. W. Montroll, "Statistical Mechanics of Nearest Neighbor Systems," *Journal of Chemical Physics* **9** (1941): 706-721; E. W. Montroll, R. B. Potts, and J. C. Ward, "Correlations and Spontaneous Magnetization of the Two-Dimensional Ising Model," *Journal of Mathematical Physics* **4** (1963): 308-322; E. H. Lieb and D. C. Mattis, *Mathematical Physics in One Dimension* (New York: Academic Press, 1966), a note on p. 469, in an article by Lieb, Schultz, and Mattis, p. 410, provides the history of the Jordan-Wigner transformation); E. H. Lieb and B. Simon, "Thomas-Fermi Theory of Atoms, Molecules and Solids," *Advances in Mathematics* **23** (1977): 22-116; P. Kastelyn, "Dimer Statistics and Phase Transitions," *Journal of Mathematical Physics* **4** (1963): 281-293; B. van der Waerden, "Die lange Reichweite der regelmassigen Atomordnung," *Zeitschrift fur Physik* **118** (1941/1942): 473-488; R. Peierls, "On Ising's Model of Ferromagnetism," *Proceedings of the Cambridge Philosophical Society* **32** (1936): 477-481; B. Widom, "Equation of State in the Neighborhood of the Critical Point," *Journal of Chemical Physics* **41** (1965): 3898-3905